**Economic-Driven Adaptive Traffic Signal Control**

**Shan Jiang**
Ph.D. Candidate
Department of Industrial and Systems Engineering
Rutgers University, New Brunswick, NJ 08854
Email: sj576@rutgers.edu

**Yufei Huang**
Ph.D. Student
Department of Industrial and Systems Engineering
Rutgers University, New Brunswick, NJ 08854
Email: yh639@scarletmail.rutgers.edu

**Mohsen Jafari, Ph.D.**
Professor and Chair
Department of Industrial and Systems Engineering
Rutgers University, New Brunswick, NJ 08854
Email: jafari@rci.rutgers.edu

**Mohammad Jalayer, Ph.D.**
Assistant Professor, Dep. of Civil and Environmental Engineering
Center for Research and Education in Advanced Transportation Engineering Systems (CREATEs)
Rowan University
Glassboro, NJ 08028
Email: jalayer@rowan.edu
ORCID ID: 0000-0001-6059-3942

Word Count: 5860 words + 2 table (500 words per table) = 6360 words

*Submitted [July 31, 2020]*

**ABSTRACT**
With the emerging connected-vehicle technologies and smart roads, the need for intelligent adaptive traffic signal controls is more than ever before. This paper proposes a novel Economic-driven Adaptive Traffic Signal Control (eATSC) model with a hyper control variable – interest rate defined in economics for traffic signal control at signalized intersections. The eATSC uses a continuous compounding function that captures both the total number of vehicles and the accumulated waiting time of each vehicle to compute penalties for different directions. The computed penalties grow with waiting time and is used for signal control decisions. Each intersection is assigned two intelligent agents adjusting interest rate and signal length for different directions according to the traffic patterns, respectively. The problem is formulated as a Markov Decision Process (MDP) problem to reduce congestions, and a two-agent Double Dueling Deep Q Network (DDDQN) is utilized to solve the problem. Under the optimal policy, the agents can select the optimal interest rates and signal time to minimize the likelihood of traffic congestions. To evaluate the superiority of our method, a VISSIM simulation model with classic four-leg signalized intersections is developed. The results indicate that the proposed model is adequately able to maintain healthy traffic flow at the intersection.
***Keywords*:** Adaptive Traffic Signal Control, Double Dueling Deep Q Network, Markov Decision Process, Reinforcement Learning

**INTRODUCTION**

Many studies have shown that adaptive signal control (ASC) improves traffic performance, such as emissions, travel time, and fuel consumption by at least 10% [1]. If the system is under saturated conditions and with extremely outdated signal timing and, the improvement can be 50% or more. However, most of the currently deployed traffic signal systems have not utilized ASC. In the United States alone, people must collectively wait 296 million hours every year, averaging one hour per person due to traffic control systems [2]. These delays negatively impact the economy and environment. There are already some adaptive solutions to the traffic signal scheduling problem in the literature [3–5]. Theoretically speaking, the optimality of these solutions is hard to establish, and computational requirements are excessive under real-life scenarios [6]. However, significant improvements in traffic congestion have been reported in many of the real-life implementations [7, 8].

In a companion work, we are laying down the foundation of an economic-based model for traffic signal control. Similar to the real economy where the prime interest rate set by the Federal Reserve Bank regulates the flow of money, we use interest rates at an intersection to control its traffic flow. For a network of roadways and intersections, there could be multiple prime interest rates prescribed by an authority at the network level. The discussion on how this works at the network level is left for a future article. In this paper, we developed a new adaptive traffic signal control using the aforementioned economic model, hereafter, to be referred to by Economic-driven Adaptive Traffic Signal Control (eATSC). Methodologically, we improved an existing approach [9] by considering the accumulated waiting time of each vehicle at an intersection for signal phase sequencing and timing. Similar to continuous compounding formula that calculates the future value of the investment in economics, the "interest rate" is introduced to measure the impact of the waiting time in a penalty (virtual currency) incurred by the intersection for keeping vehicles waiting at the traffic light. Notably, it is innovative to take these measures in defining virtual financial transactions between vehicles and the intersection control agent for signal scheduling. We assume that the intersection control agent uses a prime rate (assigned by the roadway network controller) and adjusts the actual rates for its different directions according to dynamically changed traffic conditions. The longer a vehicle waits at the intersection, the more penalty is incurred. A smart controller will find proper control strategies to reduce congestions by weighing interest rates, traffic delays, and the number of vehicles in each queue at the intersection. The interaction between road users and the controller will gradually make the controller smart enough to set the right traffic signal times and interest rates.

In recent years, the idea of using Reinforcement Learning in traffic signal control has received attention from many researchers. This type of learning process presents an unsupervised way to solve the problem where patterns are dynamically changing. Unlike supervised learning methods, it utilizes a reward signal, with no explicit mapping from input to target data, but instead aims at maximizing the reward it receives, e.g., minimum waiting time or the number of vehicles in queue. This paper develops a strategy to reduce the congestions using Double Dueling Deep Q Network and eATSC, which is generic and can be extended to various types of intersections.

**LITERATURE REVIEW**

Traffic signals along arterials and networks coordinate to provide progression to the major through movements. Traditional signal systems use fixed timing plans to prepare offline control based on historical traffic flow data. Fixed-time plans, however, cannot deal with the variability of traffic patterns throughout the day, and they become inefficient due to the traffic growth and changes in traffic patterns [10]. There are also signal systems that employ nonfixed-time plans to account for the variability of traffic flow. These plans are based on traffic volume and road occupancy from detectors located in key locations of the roadway. The system operator may also override the fixed timings based on real-time surveillance data [11].

Unlike the traditional signal timing process, "On-line" control systems update the timing plans in real-time based on detectors data. There are two major categories in this strategy: update of traffic timing plans that adjusts the signal configurations while keeping a common cycle length; and continuous optimization of signal timings at each intersection over a short time interval for Adaptive control policies. Adaptive Traffic Control System relies on accurate and fast detection of the current conditions in real-time to allow an effective response to any changes in the current traffic situation [12].

In recent articles [13–14], the authors proposed image-based deep reinforcement-learning algorithms that automatically extract all features (machine-crafted features) useful as traffic states for adaptive traffic signal control from and learns the optimal traffic signal control policy. To improve the algorithm stability and reduce the overfitting, the paper [14] also adopts Target Network, Dueling Network, and Prioritized Experience Replay. These works can adaptively adjust traffic light durations according to traffic states from raw traffic data at the intersection but adopting fixed signal sequence. There are significant gaps between theoretical advances and real-world practices in traffic control. The following key points are noted:

1. There is no study that brings economic concepts to traffic signal controls. Considering interest rates for traffic signal control strengthens the interpretability of the produced solutions.
2. Current RL-based adaptive signal control approaches either cyclically switch the signals or randomly change the signals, which loses generality.
3. Signal controls in real-time are now based on flow and occupancy observations at fixed times and locations, which may not be appropriate for all traffic conditions.

Compared to current related works, we make contributions in the following folds:

1. An economic-driven Adaptive Traffic Signal Control (eATSC) model that captures both number of vehicles and accumulated waiting times in queue is proposed for traffic signal control. By virtue of only a few control variables (interest rate and signal length), this work attempts to simplify control while adopting an artificial intelligence (AI) approach.
2. Our work allows agents to change both signal sequence and green signal times according to dynamic traffic conditions, which provides flexible ways to control the signal.
3. A two-agent Double Dueling DQN model is utilized to optimize the traffic flow by changing signal sequences and green signal durations collectively.

**METHOD AND DATA**

In this study, we considered a four-leg intersection with general configuration, as shown in Figure 1. The intersection has twelve lanes (d1-d12), and each lane represents a single direction. To be specific, the Eastbound has three lanes (d1, d2, and d3), where d1 is dedicated to the left-turn movements, d2 is to straight movements, and d3 is to right-turn movements. We assume that the maximum queue size for all directions is the same. In the following sections, we first introduced and formulated the virtual financial transactions between vehicles and the traffic signal agent, and then modeled the signal control as the Markov Decision Process.

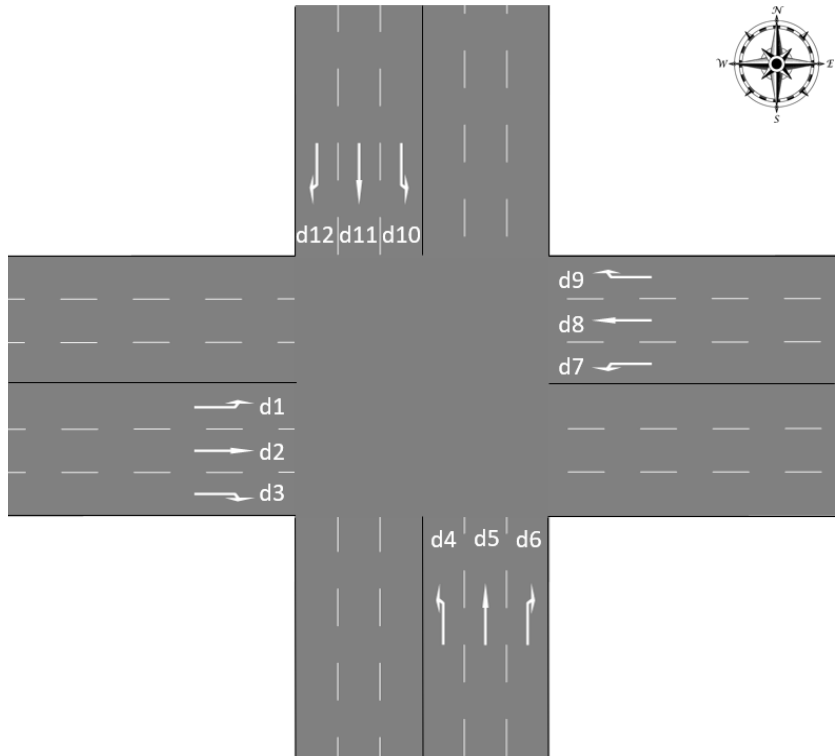

**Figure 1.  The geometry of the study intersection.**

**Economic Adaptive Traffic Signal Control Model**
     The control strategy developed here uses some of the basic elements of Organizing Traffic Light control  (SOTL) and combines these with the way interest rate and financial loans work in the economics domain. In economy, interest rate defines the time value of money (loan) transacted between a borrower and a lender. It is an exogenous quantity, which is controlled by the market where the transaction takes place. In free-market economies, a given interest rate is usually driven by the prime interest rate, which is set by the government; and its value and variations are reflective of the current and future states of the local and global economies and major indices. The basic interest rate set at the government level trickles down differently to various parts of the economy, depending on their own local conditions and global interactions.
     We used the same basic concepts in the formulation of eATSC model. Every time that a vehicle comes to a stop at an intersection and waits for a green light, it engages in a virtual financial transaction with the intersection smart controller. Here, the intersection controller is the debtor, and vehicles waiting for traffic signals are the creditors. The time value of the initial virtual money compounded on a vehicle's waiting time is the payoff amount or the penalty that the intersection controller must pay to the stopped vehicle. The intersection control will want to minimize its total payoff amount over time. The prime interest rate used by the intersection is dictated by the roadway network and depends on how this intersection performs concerning the overall network. For the purpose of this paper, we did not cover the network aspects of interest calculations and instead focus on a single intersection with interest rates for various directions controlled locally.
     First, we introduced the grouping policy for directions. Depending on traffic signal phasing and engineering design of the intersection, non-conflicting directions can be grouped into the same queue. Figure 2 demonstrates the conflicting and non-conflicting groupings.

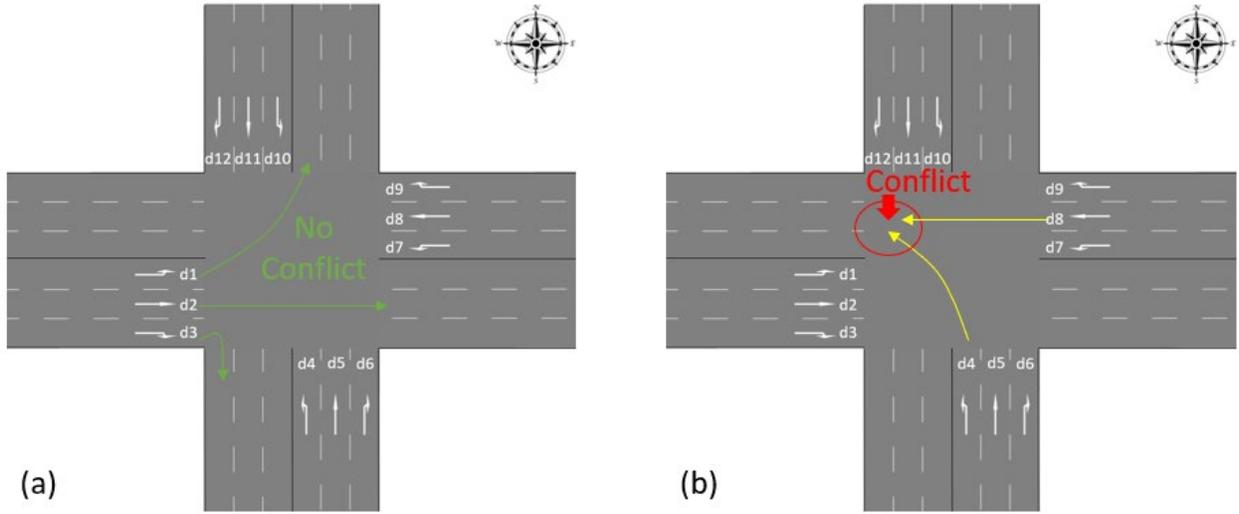

**Figure 2. (a) non-conflicting grouping, (b) conflicting grouping.**

In Figure 2(a), traffic flows along with directions d1, d2, and d3 are conflict-free and, hence can use a single green phase. In contract, releasing vehicles from d4 and d8 at the same time will cause conflicts as they will meet up at some points. Therefore, when grouping directions into queues, conflicts shall be strictly avoided. Throughout this study, we allows one group in the following list to open at the same time: {[d1, d2, d3], [d4, d5, d6], [d7, d8, d9], [d10, d11, d12]}. Therefore, vehicles from Eastbound, Westbound, Northbound, and Southbound form their own queues respectively, naming $Q_E$, $Q_W$, $Q_N$ and $Q_S$. In these configurations, all twelve directions are grouped into four non-conflicting queues. Suppose that every time a vehicle comes to a stop at an intersection along a direction, say $d$, a financial transaction is registered between the intersection smart control agent and the vehicle. This transaction involves lending a principal amount of $P$ (in virtual dollars) to the controller by the vehicle with cumulated waiting time $W$ and an interest rate of $i$. The general penalty function is presented as:

$$f(P, W, i) = P \cdot e^{W \cdot i} \tag{1}$$

Note that different directions/queues can be assigned different interest rates depending on the time of day and traffic conditions. We define a sequence of decision points in time $\{t_0, t_1, \ldots, t_{n-1}, t_n, \ldots\}$ where $\{t_0\}$ can be set to any clock time for 24 hours. For instance, for a peak period of 6:00 AM. 9:00 AM, $t_0 = 6:00\ AM$. Total penalties for a number of vehicles in direction $d$ at the time $t_n$ is defined as:

$$\delta_{d,t_n} = \sum_{j \in \Omega_{d,t_n}} P_j \cdot e^{W_{j,t_n} \cdot i_d} \tag{2}$$

where $\Omega_{d,t_n}$ is the set of vehicles travel along with direction $d$ at $t_n$ and $i_d$ is an interest rate assigned to direction $d$. Default value of $P_j$ is set to one, but we have the choice of using different values. For instance, emergency vehicles or transit buses may have higher values than passenger cars, the same for carpool vehicles with two or more passengers. $W_{j,t_n}$ is the waiting time of vehicle $j$ at $t_n$.

A cycle is a time between two red signals, and the composition of each queue can be a mix of remaining vehicles from the last cycle and new arrivals, as shown in Figure 3.

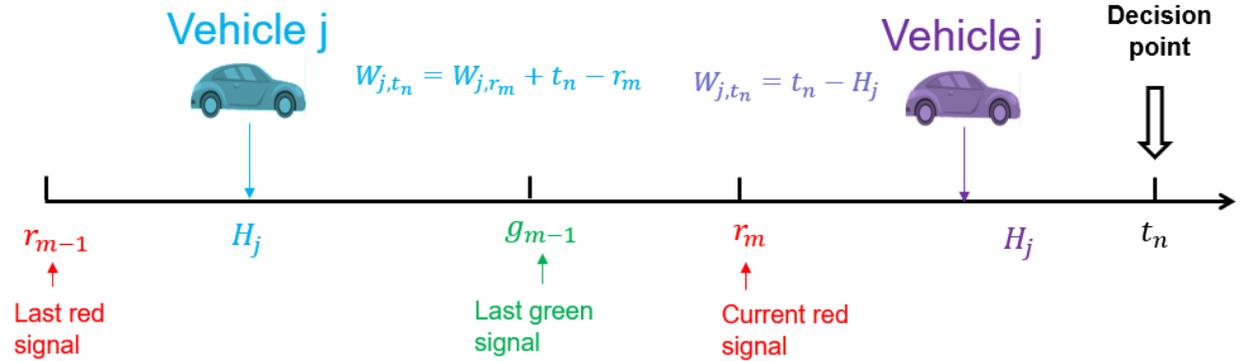

**Figure 3. Vehicle composition in the queue.**

Thus, the total waiting time of a vehicle $j$ at $t_n$ have two possibilities:

$$W_{j,t_n} = \begin{cases} W_{j,r_m} + t_n - r_m, & H_j < r_m \; j \in \Omega_{d,t_n} \\ t_n - H_j, & H_j \geq r_m \; j \in \Omega_{d,t_n} \end{cases} \quad (3)$$

where $W_{j,r_m}$ is the cumulative waiting time of vehicle $j$ that arrives in the previous cycle. $W_{j,r_m}$ is zero if vehicle $j$ arrives in the current cycle. $H_j$ is the arrival time of vehicle $j$. The total penalty for queue $q$ at $t_n$ is given by:

$$\pi_{q,t_n} = \sum_{d \in \Omega_{q,t_n}} \delta_{d,t_n} \quad (4)$$

where $\Omega_{q,t_n}$ is the set of directions in queue $q$ at $t_n$. Note that the proposed eATSC model captures both the number of vehicles waiting at the red signal for all directions $d$ and the waiting times of these vehicles, a similar concept used by SOTL. However, the treatment of waiting times through interest rates is unique. The term $e^{W_{j,t_n} \cdot i_d}$ captures continuous compounding of interest incurred over waiting times. For $i_d = 0$ the *eATSC model* reduces to the *Null model* which captures only the number of vehicles $\sum_{j \in \Omega_d} P_j$. At time $t_n$ we assign green light to all directions that belong to queue $q$ if

$$\pi^*_{q,t_n} \geq \pi_{\ell,t_n}, for\ all\ \ell \neq q \quad (5)$$

In simple terms, we look for the queue which has the highest penalty at decision time $t_n$. And the new decision can be made at the end of the current green light duration.

**Formulation of the Markov Decision Process for Signal Scheduling**

To formulate the problem, a two-step approach is proposed: the first step determines the interest rate that feeds into Equation 2 for penalty calculation, and the second step determines the green signal length according to current traffic patterns. Two agents, *Agent₁* and *Agent₂*, take care of the two steps, respectively. *Agent₁* chooses interest rates for penalty calculation in Equation 4 and switching signals for all directions based on the penalty values according to Equation 5. Both agents share the same traffic information and objective and take their own actions at the same time. Each agent is a Markov Decision Process (MDP) with $(S, A^{(l)}, R^{(l)}, P^{(l)}), l \in \{1,2\}$ representing the state space of State, Action, Reward, and Transition Probability for *Agent₁* and *Agent₂*. Definitions of each variable are as follows:

- $p^{(l)}\left(s_{t_n} \middle| s_{t_n}, a_{t_n}^{(l)}\right)$ is the transition probability for *Agent₁* from the state $s_{t_n}$ to the new state $s_{t_{n+1}}$ after taking the action $a_{t_n}^{(l)}$, where $s_{t_n}, s_{t_{n+1}} \in S$ and $a_{t_n}^{(l)} \in A^{(l)}$.
- $s_{t_n} = [s_{t_n,E}, s_{t_n,W}, s_{t_n,N}, s_{t_n,S}]$ is the state variable for traffic condition in each queue at $t_n$, where $s_{t_n,q}$ is the total number of vehicles in $q$ at $t_n$, and $q \in [E, W, N, S]$.
- $a_{t_n}^{(1)} = [i_{j,E}, i_{j,W}, i_{j,N}, i_{j,S}]$ is a list of interest rates that *Agent₁* takes at time $t_n$ for all queues.
- $a_{t_n}^{(2)} = g_j$ is the length of the green signal that *Agent₂* chooses at time $t_n$ for the next green signal.
- $r_{t_n}^{(l)}\left(s_{t_n}, a_{t_n}^{(l)}\right)$ is the reward of taking $a_{t_n}^{(l)}$ state $s_{t_n}$ at time $t_n$. It accounts for the change of queue length after taking actions,

$$r_{t_n}^{(l)}\left(s_{t_n}, a_{t_n}^{(l)}\right) = \sum_q (s_{q,t_{n-1}} - s_{q,t_n}) \quad (6)$$

We note that Equation 6 considers the difference between total vehicles in the queues before and after the two agents taking action. If the result is positive, then the performance improves, otherwise, degrades. The objective of maximizing this reward will push the agents to minimize the queue size.

In a general case, at each decision point $t_n$ (the time at $n^{th}$ signal switch), an action *a* taken at state *s* leads to a new state *s'* with probability $p(s'|s,a)$ and a reward $r(s,a)$. But it requires prior knowledge about the transition. In the absence of such knowledge, we used a two-agent Deep Q Network (DQN), which is based on sampling from experience rather than prior knowledge. The update rule was modified to take into account samples of observed data, which tend to approach transition probabilities. Different from the Offline learning requiring numerous data at one time, DQN updates the model by continuously receiving states and rewards from the environment.

The objective of DQN is to find an optimal action policy $\emptyset^*$ that maximizes the expected cumulative future reward, namely *Q-value*:

$$Q^{\emptyset^l}(s,a) = E\left[\sum_{i=0}^{\infty} \gamma^i r_{t_{n+i}}^{(l)} \middle| s_{t_n} = s, a_{t_n}^{(l)} = a, \emptyset^l\right] = \sum_{i=1}^{\infty} \gamma^{i-1} \sum_q (s_{q,t_{i-1}} - s_{q,t_i}) \quad (7)$$

where $\gamma$ is the discount factor (i.e., reward decay), which is usually between [0, 1]. When $\gamma = 1$, it equally treats future rewards, and $\gamma = 0$ takes only the current reward. In addition, we used the $\epsilon$-greedy policy to balance the exploration and exploitation during the learning process, where $\epsilon$ is within [0, 1]. It usually starts at one and decays by $\delta$ amount after every signal switch until it reaches $\epsilon_{min}$. Definition of $\epsilon(t_n)$ is as follows:

$$\epsilon(t_n) = \max(1 - \delta t_n, \epsilon_{min}) \quad (8)$$

At each decision point $t_n$, the system uniformly generates a random value and compares it to $\epsilon(t_n)$; If the value is greater, we exploit the system by taking the action that leads to the greatest *Q-value*; else, we randomly take action to explore the system. It allows the system to try different actions before knowing the consequences and eventually becomes mature to select the action that brings the highest reward. When optimal $Q^{\emptyset^{l*}}(s,a)$ for all state-action pairs are obtained, the optimal policy $\emptyset^{l*}$ for the state $s$ is simply the action $a$ that leads to the greatest $Q^{\emptyset^l}(s,a)$:

$$\emptyset^{l*}(s) = \arg\max_a Q^{\emptyset^l}(s,a) \tag{9}$$

Therefore, a recursive relationship of *Q-value* for *Agent₁* and *Agent₂* is found, known as Bellman Optimality Equation:

$$Q^{\emptyset^{l*}}(s,a) = E[r_{t_n}^{(l)} + \gamma \max_{a'} Q^{\emptyset^l}(s_{t_{n+1}}, a') | s_{t_n} = s, a_{t_n}^{(l)} = a] \tag{10}$$

This equation indicates that the optimal cumulative reward for each agent is equal to the immediate reward after taking action $a$ in traffic state $s$ plus the optimal future reward. When the optimal policy is known, the agents would know what action to take in terms of interest rate, and green signal duration is given any traffic state.

### A. DQN Models and Algorithm

Double DQN [15] and Dueling DQN [16] are both shown to reduce overestimation and thus improve the performance effectively. Double Dueling DQN increases stability during learning. For the training, we combine Double and Dueling DQN along with Prioritized Experience Replay [17] to boost the learning process. The learning has three steps, and the training process is the same for each of the two agents. Figure 6 shows the detailed learning process.

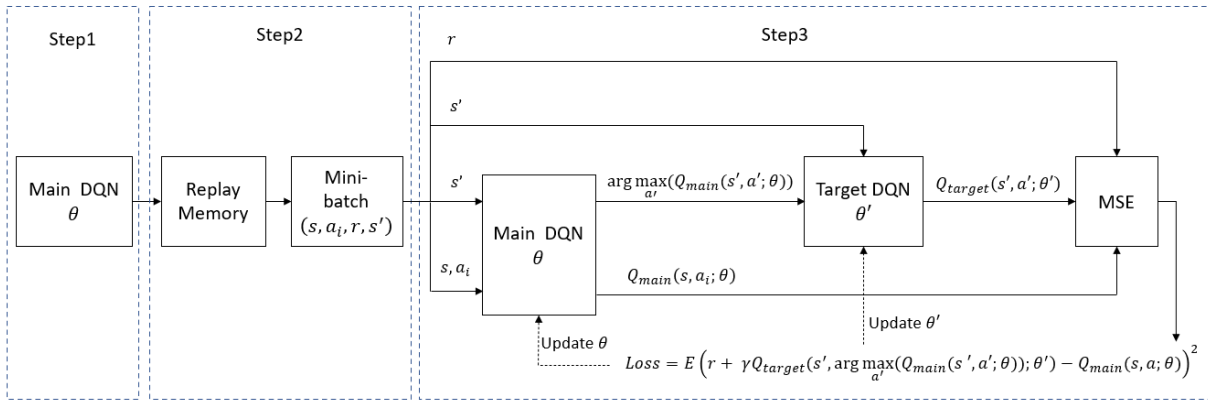

**Figure 6. The structure of Double DQN.**

In step1, the current state and the tentative actions are cast into Main Network with parameter $\theta$ to choose the most rewarding action. Then, the system interacts with the simulation environment and feeds the old state $s$, action $a$, new state $s'$, and the reward $r$ received, into the memory as a form of four-tuple $(s, a, r, s')$ in step 2. The data in the replay memory are selected by the Prioritized Experience [17] to generate mini-batches and are used to update the primary neural network's parameters. In step 3, Target Network with parameter $\theta'$ is introduced and interacts with Main Network to update the parameters by the loss function.

The main DQN with parameter $\theta$ and Target DQN with parameter $\theta'$ have the same structure, but the parameters are updated at different rates. The update rate in Main DQN is designed to be faster than Target DQN to eliminate the maximization bias. In our work, we update the Target DQN every 40 iterations. According to Equation 10, the agent tends to choose the action that leads to the highest *Q-value*; But how to make sure that the best action for the next state is the action with the highest *Q-value*? The accuracy of *Q-value* greatly depends on what action is tried and what neighboring states is explored. As a result, there is not enough information about the best action to take at the beginning of the training. Therefore, taking the maximum *Q-value* as the best action to take can lead to overestimation. If non-optimal actions are regularly given a higher *Q-value* than the optimal best action, the learning will be biased. The solution is to use two networks, main DQN and target DQN to decouple the action selection from the target *Q* value generation. The target Q value is generated by the target network and the action is generated from the main network. Similar to (7), we first define the target Q value for each agent,

$$Q_{target}(s,a) = r + \gamma Q_{target}(s', \arg\max_{a'}(Q_{main}(s',a';\theta)), \theta') \tag{11}$$

where $\theta$ and $\theta'$ denote the parameters in the Main DQN and Target DQN, respectively. To update $\theta'$ in the Target Network, we calculate the Temporal Difference (TD) Error between the Target Network and Main Network:

$$TD = |Q_{target}(s,a) - Q_{main}(s,a;\theta)| \tag{12}$$

The Target Network is then updated by the Mean Square Error (MSE) of the TD error where Proritized Experience Replay is introduced:

$$MSE = \sum_s P_s \cdot TD^2 \tag{13}$$

Here, $P_s = \frac{p_s^\alpha}{\sum_k p_k^\alpha}$ denotes the probability of getting state *s* in the training mini-batch and $p_s = \frac{1}{rank(s)}$. Rank is the location in replay memory sorted by descending order of TD error, and $\alpha$ determines how much prioritization is used. When $\alpha$ is 0, it is uniform random sampling. When $\alpha$ is 1, it only selects the experiences with the highest priorities. Prioritization also introduces bias into the system but can be corrected by reducing the weights of the often-seen samples:

$$w_s = \left(\frac{1}{M} \cdot \frac{1}{P_s}\right)^\beta \tag{14}$$

where M is the memory size, $\beta$ is linearly annealed towards 1 through the training process. The role of $\beta$ is to control how much these importance sampling weights affect learning. In practice, when $\beta$ equal to 1, the important sampling weights fully compensate for the non-uniform $P_s$.

Since *Q-value* corresponds to how good it is to be at a state and taking action at that state, we decomposed the *Q-value* as the sum of value at the current state $V(s)$ and each action's advantage compared to other actions $A(s,a)$. It is the so-called Dueling DQN [16]. The value of a state $V(s)$ denotes the overall expected rewards by taking probabilistic actions in the future steps. The advantage corresponds to every action, which is defined by $A(s,a)$. Both values are implicitly calculated by Neural Network. Figure 7 show the structure of Dueling DQN model for each agent.

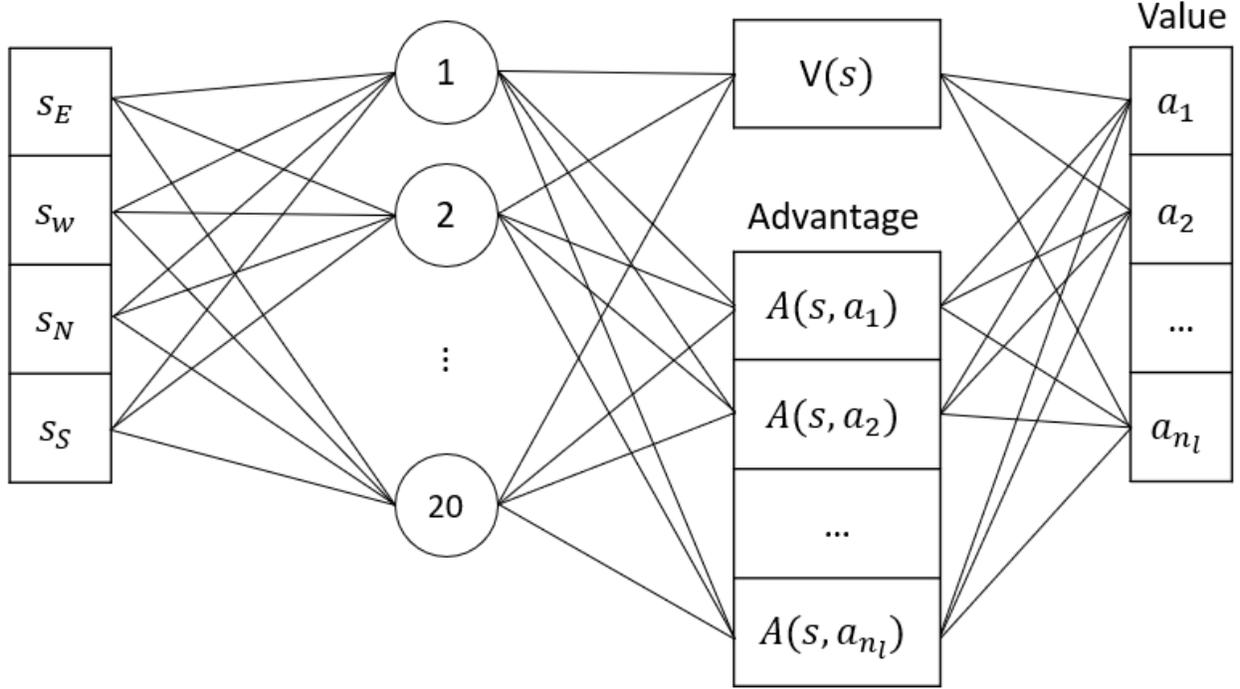

**Figure 7.  The structure of Dueling DQN**

The states of the four queues are inputs for the network, and the outputs are action values. The length of the action array and advantage array depends on the size of the action space. As described previously, the number of *inputs* is 4, and the number of outputs is 9 and 6, respectively. Besides, we assumed each Dueling DQN model has 20 nodes in the hidden layer. The pseudocode of the two-agent Double Dueling DQN with prioritized experience is shown in Algorithm 1.

**Algorithm 1: Two-agent DQN Training**

1. Initialized replay memory and DQN network
2. While $episode \leq max\_episode$ do
3.     Initialize VISSIM, $T \leftarrow 40, t \leftarrow 0, n \leftarrow 0$
4.     While $t \leq max\_simulation\_time$ do
5.         $g \leftarrow 0$
6.         $(s_{t_n}, a_{t_n}^{(1)}) \rightarrow q_{t_n}$
7.         $(s_{t_n}, a_{t_n}^{(2)}) \rightarrow g_j$
8.         While $g \leq g_j$ do
9.             $g \leftarrow g + 1,\ t \leftarrow t + 1$
10.         $(s_{t_n}, a_{t_n}^{(l)}) \rightarrow r_{t_n}$
11.         $\left(s_{t_n}, a_{t_n}^{(l)}, r_{t_n}, s_{t_{n+1}}\right) \rightarrow$ memory
12.         If memory full
13.             Train two-agent DDDQN$\rightarrow \theta^{(1)}, \theta^{(2)}$

14.         Update $\theta^{(1)'} \leftarrow \theta^{(1)}$, $\theta^{(2)'} \leftarrow \theta^{(2)}$ every T iterations
15.         $\epsilon = \max(1 - \delta, \epsilon_{min})$
16.    $n \leftarrow n + 1$
17.   Break if failure detected

## SIMULATION EXPERIMENTS

In this section, we first examined the properties of eATSC model by using static interest rates and length of signal *durations*. The work continues by showing the result of running a two-agent DQN model with variable interest rates and signal durations. The last part demonstrates the implementation of a series of pre-trained DQN control models in dynamic traffic conditions.

### Exploration of eATSC Model

Equation 1 through 5 demonstrate how to control traffic signals by interest rate. Through many experiments carried out so far, interest rates are range variables rather than single point variables. Next, we demonstrate this idea with some numerical experiments. Let us assume that we have two queues and the number of vehicles in these queues is *m* and *n*. The difference in penalty between the two queues is given by

$$D = \pi_{q_1, t_n} - \pi_{q_2, t_n} \tag{11}$$

If interest rates of two queues result in $D > 0$, then Queue 1 will be receiving the next green signal; otherwise, Queue 2. In a more general case, suppose there are two queues $Q_E$ and $Q_N$, where $Q_E$ has only one vehicle with 10 seconds waiting time, and $Q_N$ has two vehicles, and their waiting times are 8 and 9 seconds, respectively. We also assumed every vehicle carries the same principal values of one, and interest rates of directions in the same queue are identical. Table 1 lists the signal decisions for three different interest rates on the same traffic conditions.

**TABLE 1 Numerical Examples of eATSC Model**

| Case No | Parameter | $Q_E$ | $Q_N$ |
|---|---|---|---|
| 1 | Interest rate (*i*) | 0.1 | 0.1 |
|  | Penalty | 2.72 | **4.68** |
|  | Difference (*D*) | -1.96 | |
|  | Signal Decision | $Q_N$ | |
| 2 | Interest rate (*i*) | 0.5 | 0.5 |
|  | Penalty | **148.41** | 144.62 |
|  | Difference (*D*) | 3.79 | |
|  | Signal Decision | $Q_E$ | |
| 3 | Interest rate (*i*) | 0 | 0 |
|  | Penalty | 1 | **2** |
|  | Difference (*D*) | -1 | |
|  | Signal Decision | $Q_N$ | |

Since the exponential term delivers a non-linear relationship between interest rates and penalties, signal decisions in Table 1 are very different. Case 1 and 3 suggest $Q_N$ get the green signal while in example 2, $Q_E$. Clearly, there is a boundary interest rate that distinguishes the decision areas for $Q_E$ and $Q_N$. Figure 8 shows such a decision boundary for this example.

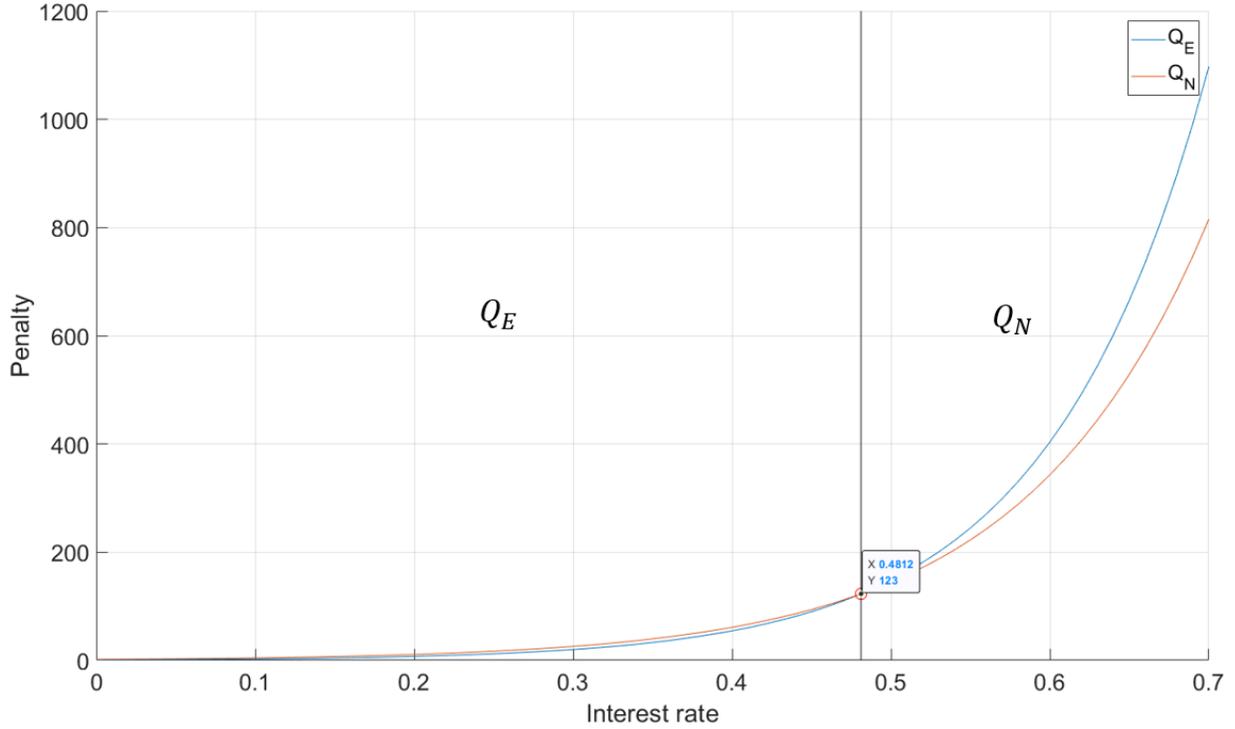

**Figure 8. Decision boundary for eATSC example.**

Given the queue size and waiting time, the boundary line is $i = 0.4812$. When interest rate to the left side of the boundary, $Q_2$ receives a green signal; otherwise, $Q_1$. The boundary suggests the necessity of discretizing the interest rate space into, and here we selected low, medium, and high for our experiments. In the sequel, we discretized the interest rates to examine their impacts on the performance of eATSC model with high/low traffic flows. The performance of the Null model also was examined. To be noted, $Q_E$ and $Q_W$ have the same traffic flow distribution (denoted as $v_{EW}$), interest rate (denoted as $i_{EW}$) and signal time (denoted as $g_{EW}$). The same assumption was made for $Q_N$ and $Q_S$. Signal times can be 10s to 60s with 10s as the interval. Traffic flows (vehicles/hour) are assumed to follow normal distributions, and interest rates can be set to Null, or a combo from the interest rate set {0.1, 0.2, 0.3}.

**Comparison of different control policies**

In this section, we compared the results of three control policies: non-fixed time and sequence, fixed time and sequence, and fixed time and non-fixed sequence, in an unbalanced traffic scenario where the traffic volume is normally distributed: $Q_E$, $Q_W \sim N(300,30)$, $Q_N$, $Q_S \sim N(600,60)$. In the first policy, both signal time and sequence are subject to change, and we used DDDQN to train the agents for interest rates and signal times selection in each cycle. The second policy adopts fixed signal time and cyclically switches the signal sequence. The last policy uses fixed signal time but the switch of signals based on Equation 5. Table 2 lists the parameters for DDDQN, and Figure 9 shows the results of each policy.

**TABLE 2 Parameters Used in Training**

| Parameter | *Policy 1* |
|---|---|
| Maximum episode | 150 |
| Maximum simulation seconds | 10,000 |
| Memory size | 500 |
| Mini-batch size | 32 |
| Main-net update frequency | 40 |
| Greedy decrement | 0.008 |
| Min greedy | 0.02 |
| Reward decay | 0.95 |
| Learning rate | 0.005 |
| Prioritization alpha | 0.6 |
| Prioritization beta | 0.4 |

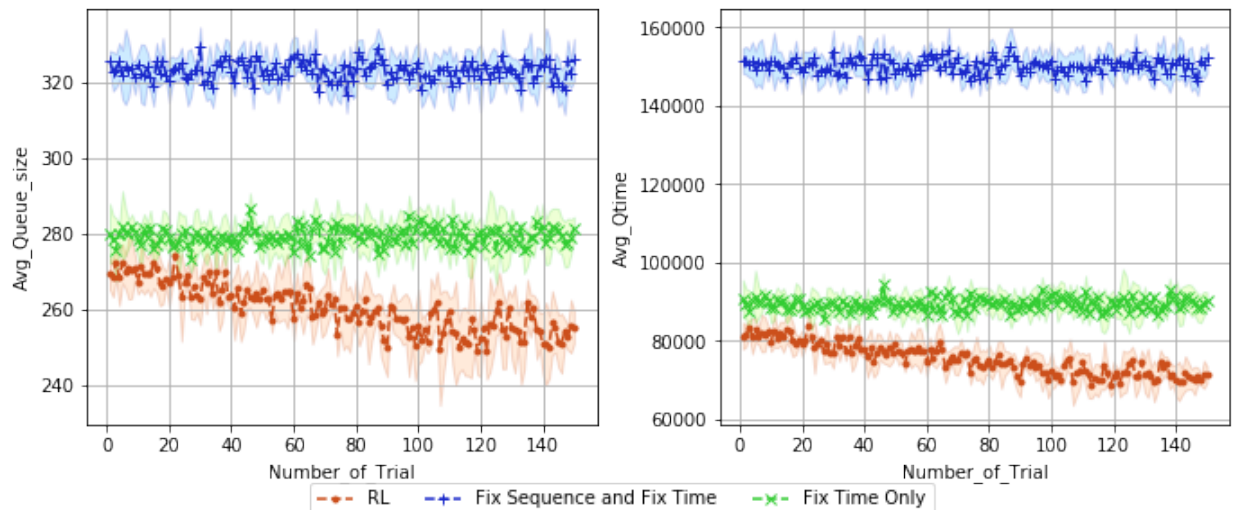

**Figure 9. Performance of three control policies**

All policies ran five replications, and both mean and standard deviation were plotted in the graph. The reinforcement learning-based method outperforms the other two in terms of queue size and waiting time, and it is evolving. In the beginning, the agents randomly sample on actions to explore the reward, queue size, and waiting time it receives. Gradually, the agents gather enough experience to take good actions for higher rewards, smaller queue sizes, and shorter waiting times. An optimal control policy appears after the system converges. It is also seen that using penalty as the switching criterion for fixed signal time outperform fixed sequence and fixed signal time. The results of this experiment show the superiority of our method and will lay a foundation for the following study.

**Simulation Results**

Different from Figure 9 that each trial can run to the maximum simulation time, in this study, we will introduce an interesting and yet strict rule: system failure, to compare the Null model (zero interest rate) and eATSC model (non-zero interest rate). *System failures* happen when any queue reaches its maximum capacity. Such failure will lead to the immediate end of the current simulation trial and the restart of a new one. In the experiment, we monitored how long the non-failure time was as well as the number of vehicles in the queue and their waiting times.

Performance metrics were recorded at the end of each cycle. Besides, the maximum capacity of all queues for holding vehicle was designed to be 110 vehicles.

We ran the simulation in an unbalanced traffic scenario where the traffic volume in $Q_E$, $Q_W \sim N(350,35)$ and $Q_N$, $Q_S \sim N(700,70)$. All essential parameters remained similar to Table 2, but the total number of trials was shortened to 120, and each trial had 6,000 simulation seconds for fast computation. Figure 10 (a) - (d) compare the results of the eATSC ($i \neq 0$) and the Null model ($i = 0$).

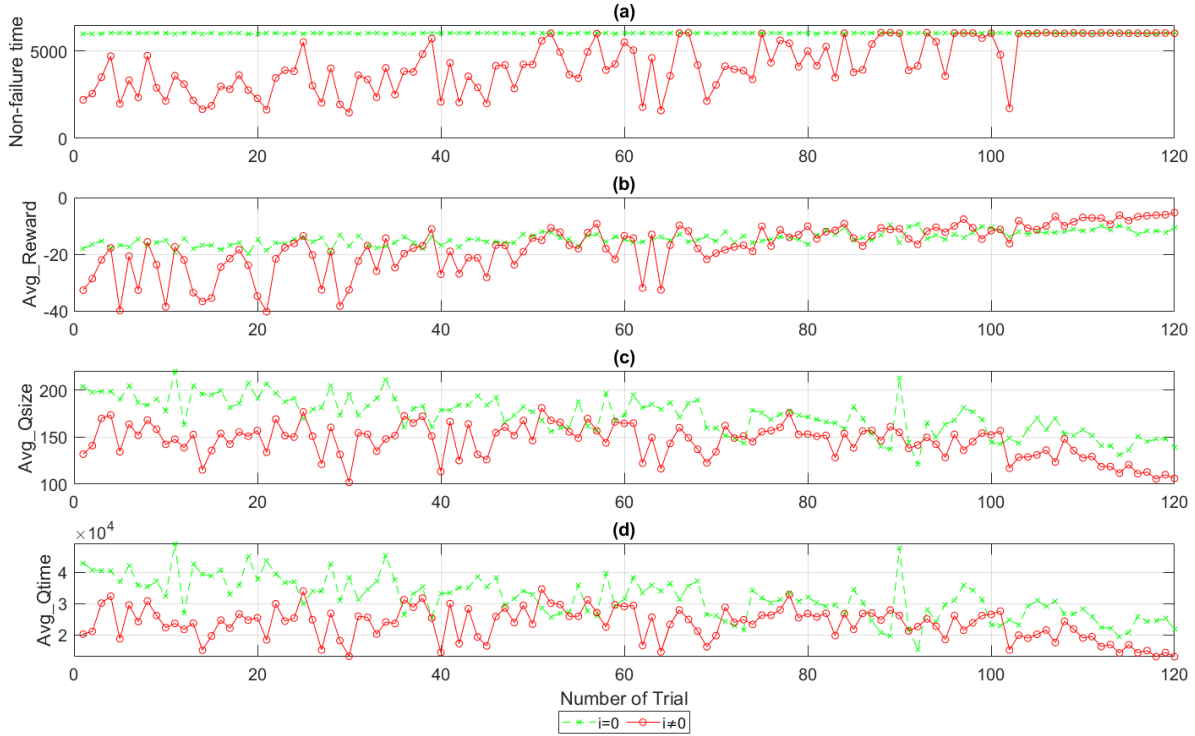

**Figure 10. System performance in terms of (a) non-failure time (b) average reward ( (c) average queue size (d) average waiting time**

Each dot represents a trial and we observe convergences due to the greedy decrement after each trial in Table 4. Challenges here are brought by system failures that occurred in the early exploration period. Early fluctuation in Figure 10(a) reflects such a challenge. The fluctuation becomes subtle and performance continuously improves and converges in the last few trials, meaning the agents acquire enough knowledge to control the traffic signal. In steady-state (e.g. trial 120), the eATSC model is found to outperform the Null model since its average queue size and waiting time are lower. To confirm the findings, we took a closer look at the time-series performance of the last trial over 6,000 simulation seconds for both models. Figure 11 (a) - (b) compare the system performance between the eATSC and the Null model in the steady-state. Both the average queue size and waiting time of eATSC model outperform the Null model. Intuitively, less busy queues would suffer more from the Null model since busy queues always outnumber them. The introduction of interest rates would theoretically solve this bias since it takes waiting times into consideration. Therefore, less busy queues would be the key to the improvement of the whole network. Figure 11 and 12 examine the performance of $Q_E$ and $Q_N$ in two models respectively in steady-state.

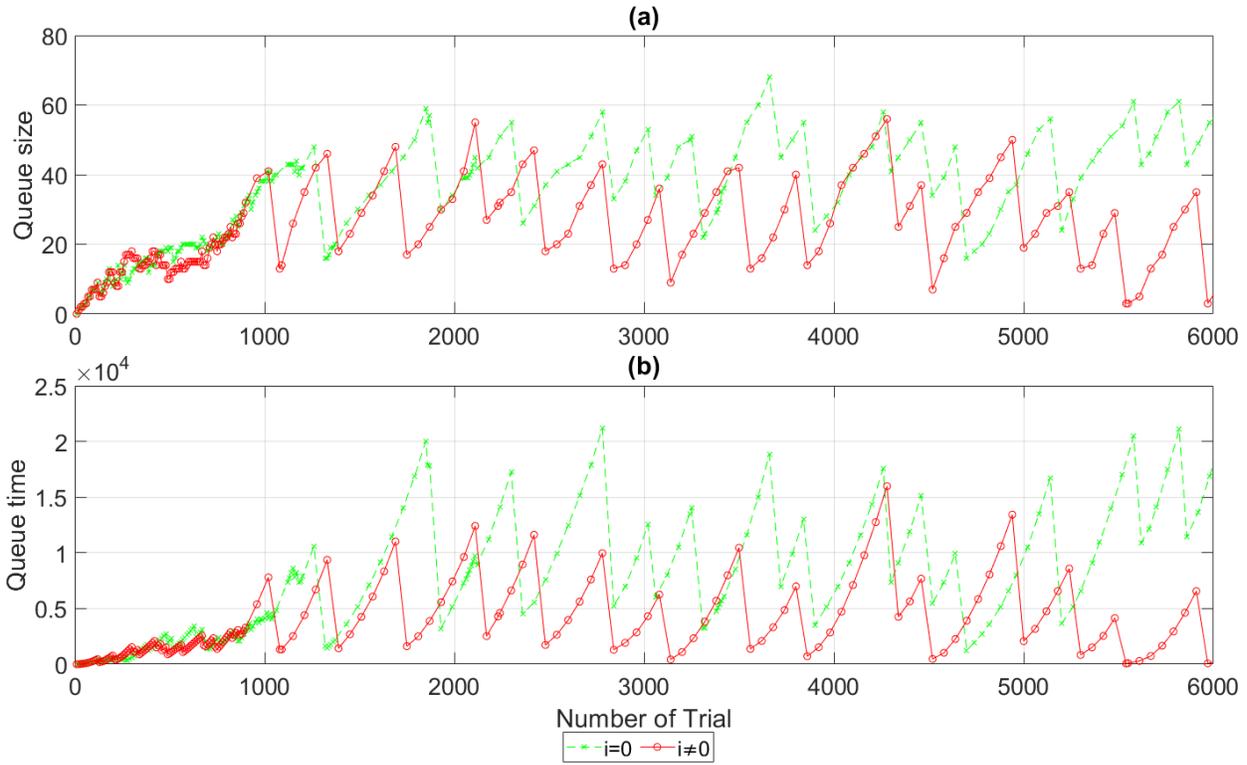

**Figure 11 - Performance of $Q_E$ in steady-state in terms of (a) queue size (b) waiting time**

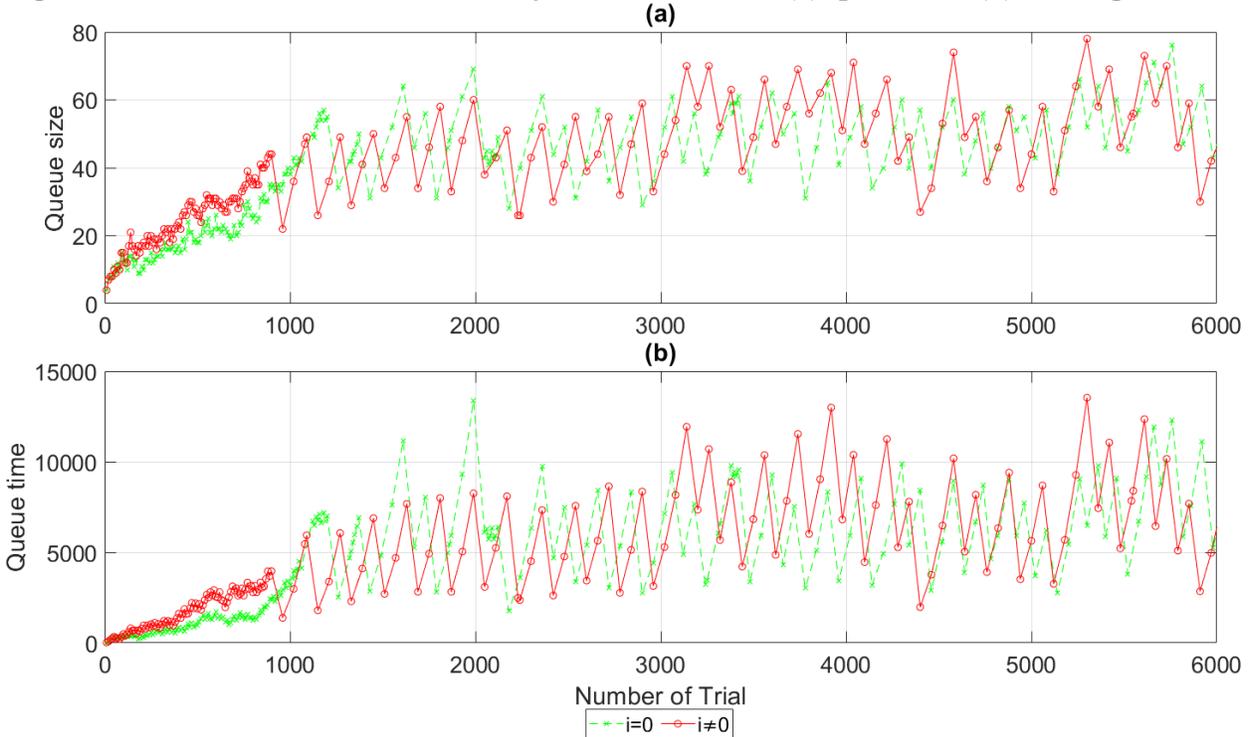

**Figure 12 - Performance of $Q_N$ in steady-state in terms of (a) queue size (b) waiting time**

According to Figure 11, the eATSC model improved the performance of lesser queues. And Figure 12 showed close performance between the eATSC and the Null model in busy queues.

By super-positioning the performance of all queues, the eATSC performed better. It was revealed that the trained DQN model could handle traffic variations within the identified arrival distribution. However, if the distribution is out of scope, a single trained model can no longer fit. We can either adopt the transfer learning [18] to retrain the model or prepare several models for different traffic conditions.

**CONCLUSION**

In this paper, we proposed the economic-driven Adaptive Traffic Signal Control (eATSC) model for traffic signal control at the intersection. The eATSC model computes penalty that grows with vehicle waiting times according to a continuous compounding economic model, with an interest rate that varies with traffic flow. The calculated penalty is utilized for signal scheduling. It is possible to have different interest rates at an intersection corresponding to different directions. In a dynamic traffic study where both interest rates and signal times are subject to change, each intersection controller is assigned with two intelligent agents seeking to minimize the traffic delay and the number of vehicles in queue. The control problem is formulated as a Markov Decision Process (MDP) problem, and a two-agent Double Dueling DQN model with Prioritized Experience Replay is utilized to solve it. The trained agents adopt optimal control policies for signal times and interest rates as functions of traffic patterns. Under the optimal policy, the chance of congestion is minimized. We first examined the property of the eATSC model and showed the superiority of our method under different scenarios. Following this, we compared the performance of the eATSC model to the Null model that solely considers queue size and disregard the waiting time. The results demonstrated that the performance of the eATSC model is significantly better in unbalanced traffic flow due to its ability to balance the queue size and waiting time. It is worth to mention that the proposed methodology is generic and can be extended to various types of intersections.